\documentclass[aps, superscriptaddress,floats,showpacs]{revtex4-2}

\usepackage{graphicx}
\usepackage{amsmath}
\usepackage{amsfonts}
\usepackage{amssymb}
\usepackage{amsbsy}
\usepackage{color}
\usepackage{bm}

\usepackage{hyperref}
\usepackage[capitalize]{cleveref}
\usepackage{graphicx}
\hypersetup{
 colorlinks,
 citecolor=blue,
 filecolor=blue,
 linkcolor=blue,
 urlcolor=blue}

\begin{document}

\title{Bright synchrotron radiation from relativistic self-trapping of a short laser pulse in near-critical density plasma}

\author{M. G. Lobok}
\affiliation{P. N. Lebedev Physics Institute,
	Russian Academy of Science, Leninskii Prospect 53, Moscow 119991,
	Russia}
\affiliation{Center for Fundamental and Applied Research,
	Dukhov Research Institute of Automatics (VNIIA), Moscow 127055, Russia}
\author{ I. A. Andriyash}
\affiliation{Laboratoire d’Optique Appliquée, ENSTA-CNRS-Ecole Polytechnique, UMR7639, 91761 Palaiseau, France}
\author{O. E. Vais}
\affiliation{P. N. Lebedev Physics Institute,
	Russian Academy of Science, Leninskii Prospect 53, Moscow 119991,
	Russia}
\affiliation{Center for Fundamental and Applied Research,
	Dukhov Research Institute of Automatics (VNIIA), Moscow 127055, Russia}
\author{V. Malka}
\affiliation{Department of Physics and Complex Systems, Weizmann Institute of Science, Rehovot 76100, Israel}
\author{V. Yu. Bychenkov}
\affiliation{P. N. Lebedev Physics Institute,
	Russian Academy of Science, Leninskii Prospect 53, Moscow 119991,
	Russia}
\affiliation{Center for Fundamental and Applied Research,
Dukhov Research Institute of Automatics (VNIIA), Moscow 127055, Russia}

\begin{abstract}
In a dense gas plasma a short laser pulse propagates in relativistic self-trapping mode, which enables effective conversion of laser energy to the accelerated electrons. This regime sustains effective loading which maximizes the total charge of the accelerating electrons, that provides a large amount of betatron radiation. The 3D particle-in-cell simulations demonstrate how such regime triggers X-ray generation with 0.1-1~MeV photon energies, low divergence, and high brightness. It is shown that a 135~TW laser can be used to produce $3\times 10^{10}$ photons of $>10$~keV energy and a 1.2~PW laser makes it possible generating about $10^{12}$ photons in the same energy range. The laser-to-gammas energy conversion efficiency is up to $10^{-4}$ for the high-energy photons, $\sim 100$~keV, while the conversion efficiency to the entire keV-range x-rays is estimated to be a few tenths of a percent.  
\end{abstract}

\maketitle

Laser-plasma accelerator can produce high-energy and high-current electron bunches on a short distance together with bright synchrotron x-rays beams \cite{Rousse_2004, Phuoc_2006}. The specific features of the so-called \textit{betatron} source, are the small source size  (down to $\sim 1$~$\mu$m \cite{Kneip_2010}), ultrashort duration (few fs \cite{Lundh_2004}) and relatively low divergence ($\sim10$~mrad \cite{Phuoc_2006}), makes it very attractive for applications in biology, chemistry, medicine, and material science \cite{Malka_2008}. Allowing for the spectrum control and being well collimated betatron x-rays were recently used for the X-ray phase-contrast micro-tomography, proving  that such sources have reached the verge of practical applications \cite{Wenz_2015}.

Synchrotron x-ray radiation from electrons laser-plasma accelerators can be generated by two different mechanisms: laser wakefield acceleration (LWFA) and direct laser acceleration (DLA). In the first case, the electrons propagate within plasma cavity (so called "bubble" \cite{Pukhov_2002,Lu_2007}), which accelerates them by the longitudinal electrostatic field. At the same time, the particles oscillate in the induced transverse electric and magnetic fields, which transversally increase in strength. If based on multi-TW lasers, such sources typically deliver X-rays in the keV range \cite{Albert_2016, Bloom_2020}, and their brilliance is limited by the total charge of accelerated electron bunches, which doesn't exceed a few hundred pC. The divergence angle of the radiation generated in LWFA regime is about a few tens of milliradians \cite{Phuoc_2006, Corde_2013}. In the DLA regime, the particles are accelerated in the plasma channel by the laser pulse and also experience the impact of the transversal fields of the channel. Such interaction extends the radiation energy in the MeV range and higher \cite{Huang_2016}.  However the divergence of such X-rays is significantly greater, reaching the order of a radian \cite{Kneip_2008, Huang_2016}.  The increase of photon energies due to the interaction of oscillating electrons with the back of the laser pulse was predicted theoretically in  \cite{Thomas_2009}. In a near critical density plasma electrons experience strong betatron oscillations that result of both enhanced plasma and laser fields within a charged cavity that leads simultaneously to the hard photons, high brightness and sufficiently low divergence. 

Our recent work indicates the possibility to maximize the charge of the LWFA electron bunches considering a particular laser pulse propagation mode \cite{Lobok_2018, Bychenkov_2019}. This occurs when laser light experiences the self-trapping so that its diffraction divergence is balanced by the relativistic nonlinearity, while the relativistic self-focusing on the axis also does not happen and the laser beam radius is approximately preserved during its propagation over many Rayleigh lengths \cite{Bychenkov_2019}. This self-trapped structure constitutes a spatial soliton, which belongs to the class of solutions in the self-focusing theory discussed in the pioneer works \cite{Townes_1964,Akhmanov_1966}.  Another advantage of this regime, is that in the high-density plasma the ion cavity can support a very high charge of injected bunch. It has been shown that short 135~TW laser pulse is able to produce about 7~nC electron beam with an energy higher than 30~MeV when interacting with a ten percent critical density plasma \cite{Lobok_2018}. 

In this article we show, using 3D PIC simulations with a high performance relativistic electromagnetic code VSim (VORPAL, \cite{VORPAL}), that such regime is also well suited to enhance significantly the yield synchrotron radiation. Laser pulse linearly polarized along the z-axis propagates along the x-axis and has the following characteristics: the wavelength $\lambda = 2 \pi c /\omega_l =1$~$\mu$m, Gaussian temporal envelope with FWHM duration $\tau =30$~fs, and Gaussian amplitude profile of the focal spot with the FWHM size $D=2R_L =4\lambda= 4$~$\mu$m.  We consider two values of the standard normalized laser field amplitude, $a_0=eE_L/m_e\omega_l c$:  $a_0= 24$ and 72, which correspond to maximum laser pulse intensities of $\simeq 8\times 10^{20}$ and $7\times 10^{21}$~W/cm$^2$, respectively. The laser pulse was focused on the front side of a plasma target consisting of electrons and He-ions. The electron density, $n_e$, and target thickness were chosen to maximize the total charge of electrons accelerated to characteristic energy $\gtrsim 30$~MeV. The chosen parameters enable stable propagation of the laser pulse over many Rayleigh lengths and effective generation of electron bunch with high average energy and high total charge. Thus, for $a_0 = 24$, these parameters were $n_e = 0.1\,n_{c}$ and 240~$\mu$m, where $n_c$ is the critical density.  For $a_0 = 72$, we used $0.3\,n_{c}$ and 200~$\mu$m. The target thickness corresponded to almost entire laser pulse depletion in both cases. The simulations were performed with moving window technique with the spatial grid steps $0.02 \lambda \times 0.1 \lambda \times 0.1 \lambda$ in a simulation window $X \times Y \times Z = 58 \lambda \times 28 \lambda \times 28 \lambda$. The representative sample of electron tracks obtained in PIC simulations, was further used as an input source for synchrotron emission calculations. 

In the near-critical density plasma regime, the resulting light structure is in fact a mixed electromagnetic-electrostatic 3D soliton, which can propagate over ten Rayleigh lengths. The performed simulations clearly show importance of matching the self-consistent waveguide radius, $R$, to the electron plasma density, $n_e$, for the stable propagation of the light structure with given laser power as was demonstrated before in the numerical experiments, e.g. \cite{Gordienko_2005,Lu_2007,Lobok_2018},
\begin{equation} 
\!\!R \!\simeq\! \alpha\frac{c\sqrt{a_0}}{\omega_p}\!\!=\!\! \frac{\alpha c}{\omega_l}\! \sqrt{\!a_0\frac{n_c}{n_e}},\,{\rm or}\,
 R\!=\!\frac{c}{\omega_l}\sqrt{\frac{n_c}{n_e}}\!\!\left(\!\!\frac{16\alpha^4P}{P_c}\!\right)^{\!\!{1/6}}\!\!\!.\label{eq1}
\end{equation}
Here $\omega_p = 2\pi c/\lambda_{p}$ is the electron plasma frequency, $P$ is the given laser power, $P_c = 2\, (m_ec^3/r_e) \, (\omega_l^2/\omega_p^2) \simeq 17 (n_c/n_e)$~GW is the critical power for relativistic self-focusing \cite{Sun_1987}, $r_e=e^2/m_e c^2$ is the classical electron radius, and $\alpha$ is the numerical factor of the order of unity. The condition, \cref{eq1}, corresponds to the self-trapping regime of relativistic self-focusing (see, for example, \cite{Sen_2013,Kovalev_2019}), 
\begin{figure} [!ht]\centering{\includegraphics[width=1\linewidth]{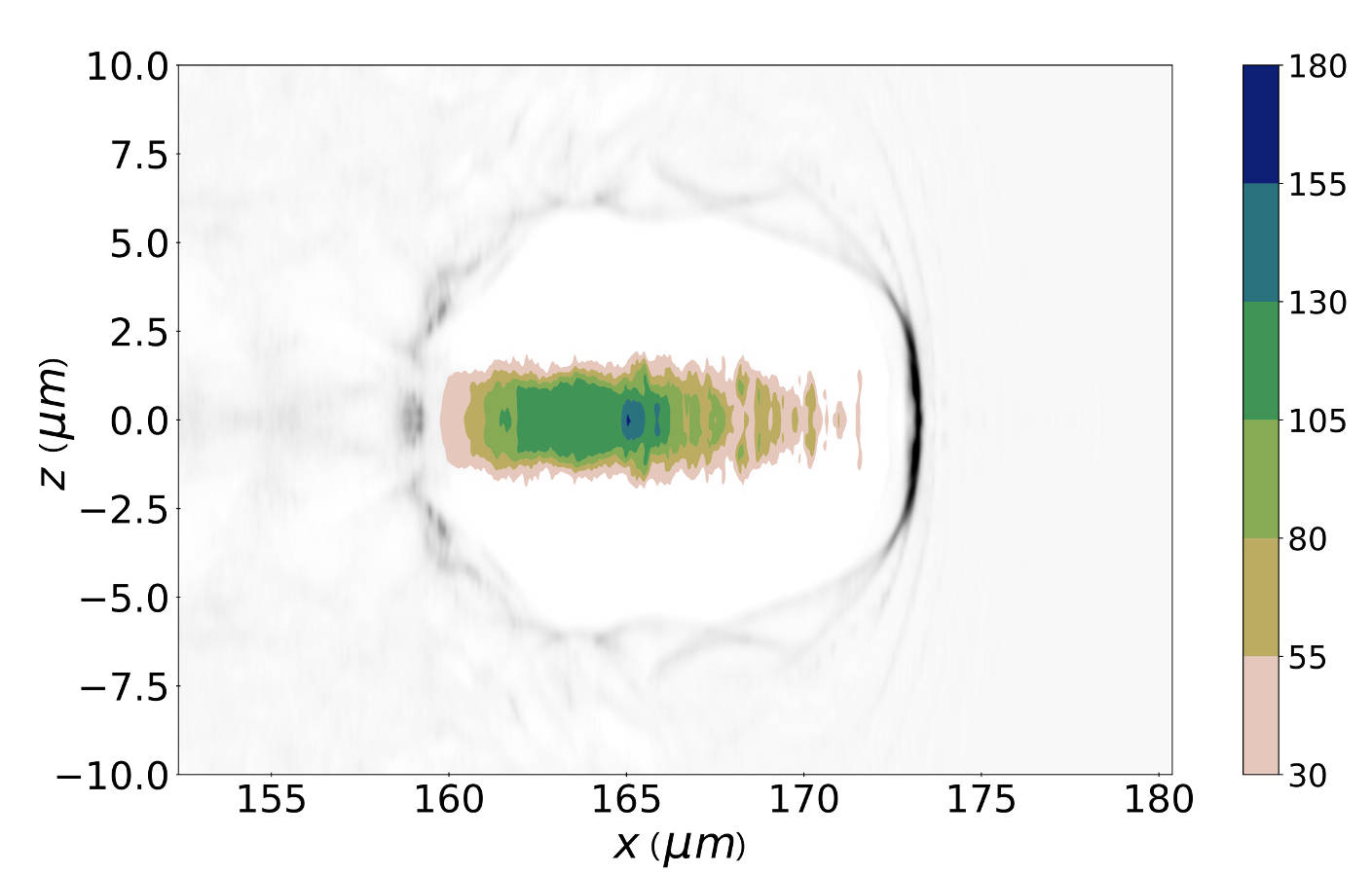}}
	\caption{The plasma cavity formed by the laser pulse propagating in self-trapping regime for $a_0 = 24$, $n_e = 0.1 n_{cr}$. The color corresponds to the average energy of the accelerated electrons. }
	\label{fig1}
\end{figure}
which occurs for the laser pulses with the focal radius at the plasma entrance being close to the self-consistent radius. The initial laser pulse radius approaching the value \cref{eq1} only weakly changes with propagation distance in accordance with relativistic self-trapping. Such propagation is very stable, since, in addition to the condition of the nonlinear optical self-consistency, the considered relation ensures both stability with regard to light filamentation and force stability as a balance between ponderomotive displacing force and focusing Coulomb force acting on electrons in the radial direction. 

Ambient electrons can flow into the traveling ion cavity from its back side similar to the self-injection in the case of a standard bubble wakefield. Then the particles are accelerated in the cavity experiencing both the high-frequency laser field and the longitudinal electrostatic plasma field. The latter field is decelerating for electrons in the front part of the cavity, and is accelerating at the back. The electrons co-propagate together with the laser pulse and electrostatic field for a rather long time, but finally laser field got etched during propagation and particles motion is becoming more and more dominated by the longitudinal electrostatic field. Pre-acceleration mechanism of the electrons could be referred to as specific kind of direct laser acceleration enabling an effective loading of high amounts of particles into accelerating electrostatic plasma field. An amplitude modulated single-cycle electrostatic field of the cavity contribute to a stochastic manner of electron acceleration similar to that observed for stimulated Raman scattering \cite{Bochkarev_2014}. Stochastic behaviour of the accelerating electrons has been demonstrated in \cite{Lobok_2018}.
\begin{figure*} [!ht]\centering{\includegraphics[width=16cm]{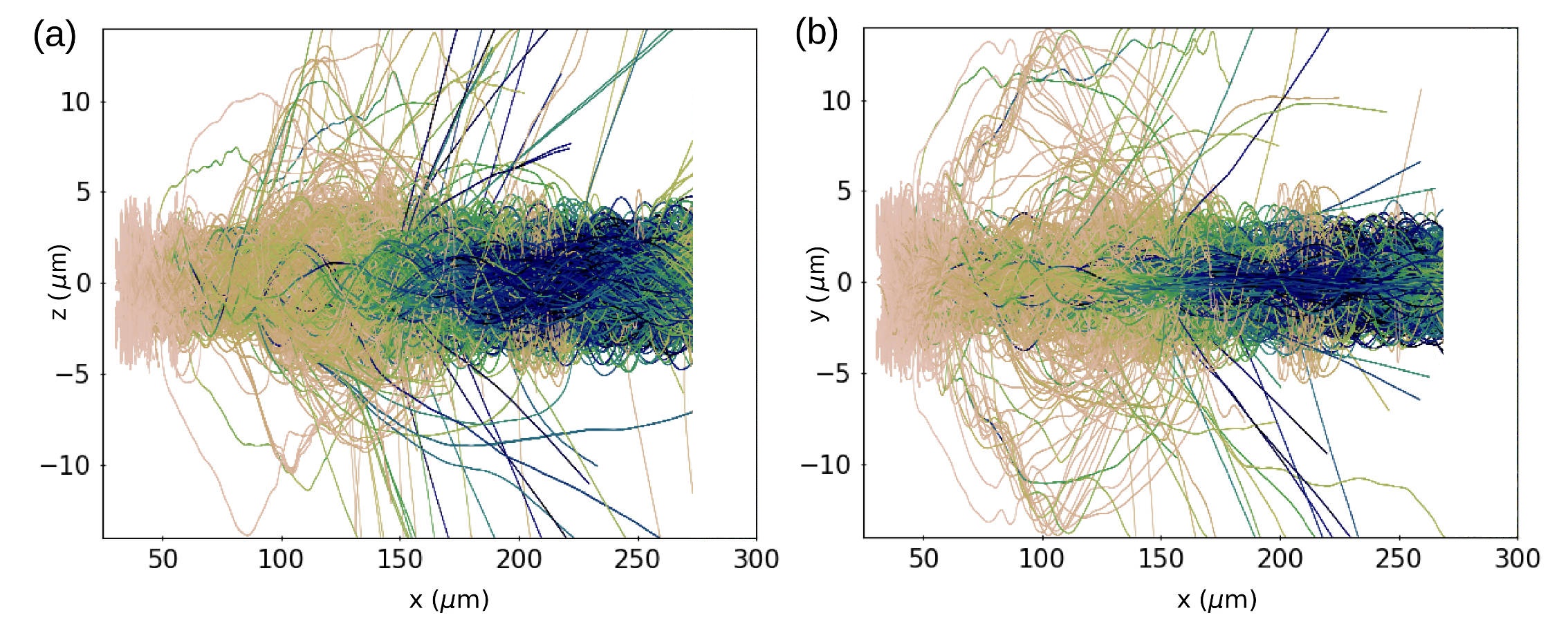}}
	\caption{Projections of the electron trajectories to the planes: polarization plane $(Oxz)$ and perpendicular one $(Oxy)$, for $a_0 = 24$. The line colors correspond to the particle energy, the dark blue refers to the maximum values.}
	\label{fig2}
\end{figure*}
The cavity formed for $a_0 = 24$ is shown in \cref{fig1} by gray color, and brown-green colors indicate distribution of the average energy of the injected electrons (with energies exceeding 30~MeV). The cavity plasma field makes it possible to accelerate even the electrons initially at rest to energies up to few hundreds of MeV, that we show in the Supplemental Materials []. The number of such energetic electrons is significantly higher than in the bubble regime. Their charge achieves about 7~nC with total energy of 0.8~J. Thus, this high-charge electron bunch is a source of high brightness betatron radiation.

In order to explore properties of the generated synchrotron radiation we have considered a sample of the randomly chosen 3000 electron trajectories, with energies in excess of 30~MeV. \Cref{fig2} shows these trajectories for a cavity formed under the following laser-plasma parameters: $a_0 = 24$, $n_e = 0.1n_{cr}$. The lines are colored according to the particle energies, so that the darkest color refers to the highest energy. Analyzing betatron amplitudes and electron energies, we can separate the process of the electron acceleration to two regimes. The first stage is characterized by the stronger betatron oscillations and lower electron energies depicted in beige in \cref{fig2}, as compared with the second one, which is illustrated by dark blue. The betatron amplitudes of the electron oscillations are slightly larger in average along the laser polarization \cref{fig2}a than along the $y$-axis (\cref{fig2}b). This effect maintains even after the termination of the particle interaction with the laser pulse, that is discussed in the Supplemental Materials []. Although the laser pulse doesn't lead to the gain of the particle acceleration, it participates in the process of stochastic electron injection \cite{Lobok_2018} and helps to produce a high accelerated charge in the considered near critical density plasma.  After declining the impact of the laser pulse on the electrons, the trapped particles are accelerated forming a plateau in the particle energy spectrum (see Fig. 9 in Ref. \cite{Bychenkov_PoP_2019}). 
\begin{figure*} [!ht]\centering{\includegraphics[width=16cm]{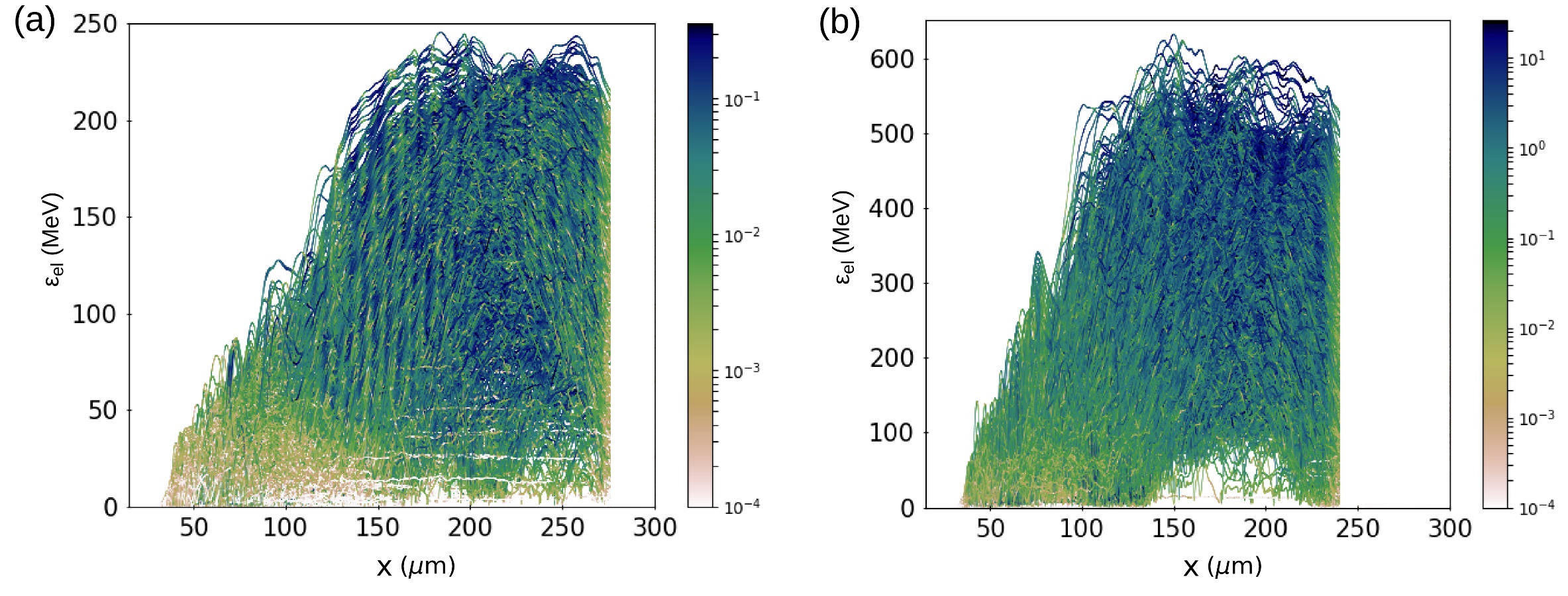}}
	\caption{Evolution of the electron energy for $a_0 = 24$ (a) and $a_0 = 72$ (b) during particle propagation trough the target. The line color refers to the instantaneous emission power radiated by the particles [keV fs$^{-1}$].  }
	\label{fig3}
\end{figure*}
\begin{figure*} [!ht]\centering{\includegraphics[width=16cm]{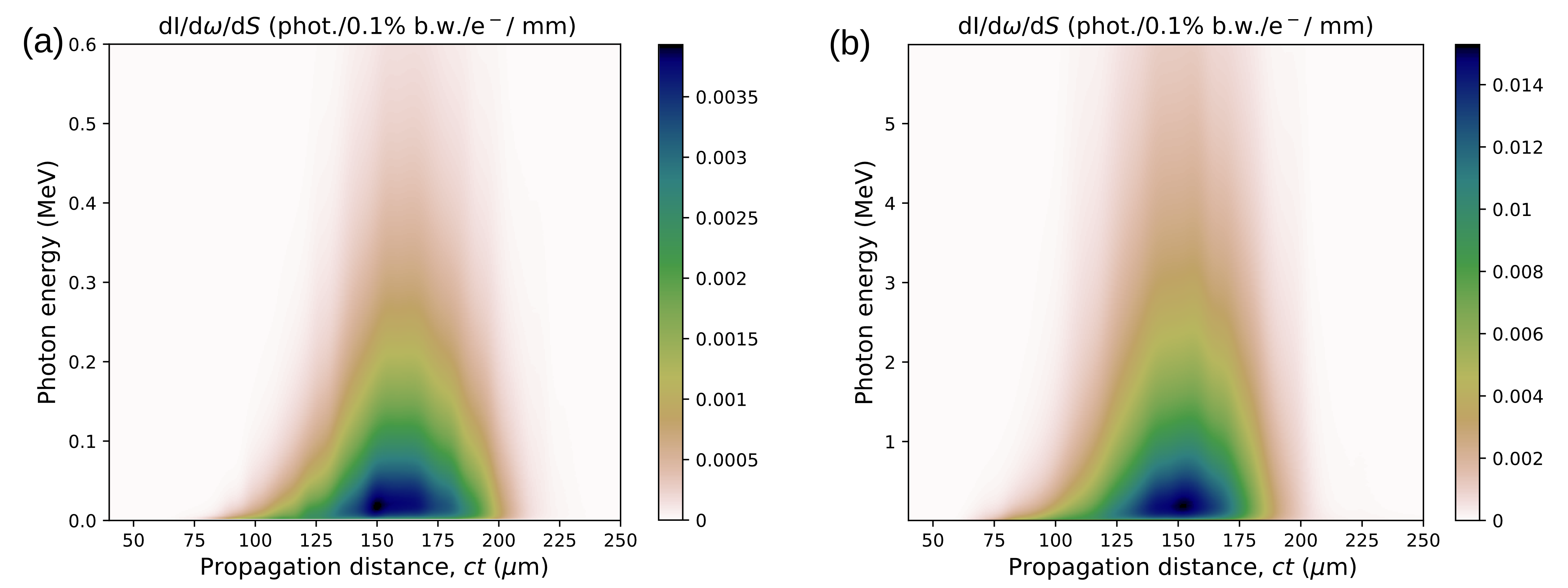}}
	\caption{Angular-integrated spectra of photons emitted per single electron along the laser propagation through the target: $a_0 = 24$ (a) and $a_0 = 72$ (b). }
	\label{fig4}
\end{figure*}
The total instantaneous radiated power can be calculated by the relativistic generalization of Larmor's formula \cite{jackson}:  
\begin{equation}
 P = \frac{2}{3}\frac{e^2}{c}\gamma^6(\boldsymbol{\dot{\beta}}^2 - [\boldsymbol{\beta} \times \boldsymbol{\dot{\beta}}]^2),
\end{equation}
where $\boldsymbol{\beta} = {\bf v}/c$ is the dimensionless velocity vector, $\gamma$ is the Lorentz factor of the electron. \Cref{fig3} shows the evolution of electron energies along the propagation distance for two different laser intensities. The line color refers to the total radiated power in the logarithmic scale. The most of the secondary radiation is emitted, when the electron achieves the highest energies. The latter happens at the second stage of the electron acceleration, although the betatron amplitudes some what decrease. 

We use 3D multiparticle tracking code \textit{Synchrad} \cite{Andriyash_2017}, which calculates angular-spectra distributions of the secondary radiation by integrating the following well-known expression:
\begin{equation}
 \!\!\frac{d^2I}{d\Omega d\omega}\!\! =\!\! \frac{e^2}{4\pi^2c}\!\left| \int^{\infty}_{-\infty}\!\!\! \frac{[{\bf n}\!\times \![({\bf n}\! -\! \boldsymbol{\beta})\!\times\!\boldsymbol{\dot{\beta}}]]}{(1 - \boldsymbol{\beta}\cdot{\bf n})^2}e^{i\omega \{t - {\bf n}\cdot{\bf r}(t)/c \}}dt\right|^2\!\!\!\!, \label{eq_I}
\end{equation}
where ${\bf r}$ is the radius-vector of the electron,  ${\bf n}$ is the unit vector of observation \cite{jackson}. After that, the amplitudes obtained for different particles are added. \Cref{fig4} shows the angular-integrated instantaneous spectra emitted along the propagation distance. The broadest high-energy wing of the spectrum is emitted at the second stage of the electron acceleration, when laser field is already weakened within a cavity region where 
dominant emitting electrons are concentrated, i.e. at the distance exceeding 100~$\mu$m. 
 
\begin{figure*}[!ht]
	\center{\includegraphics[width=16cm]{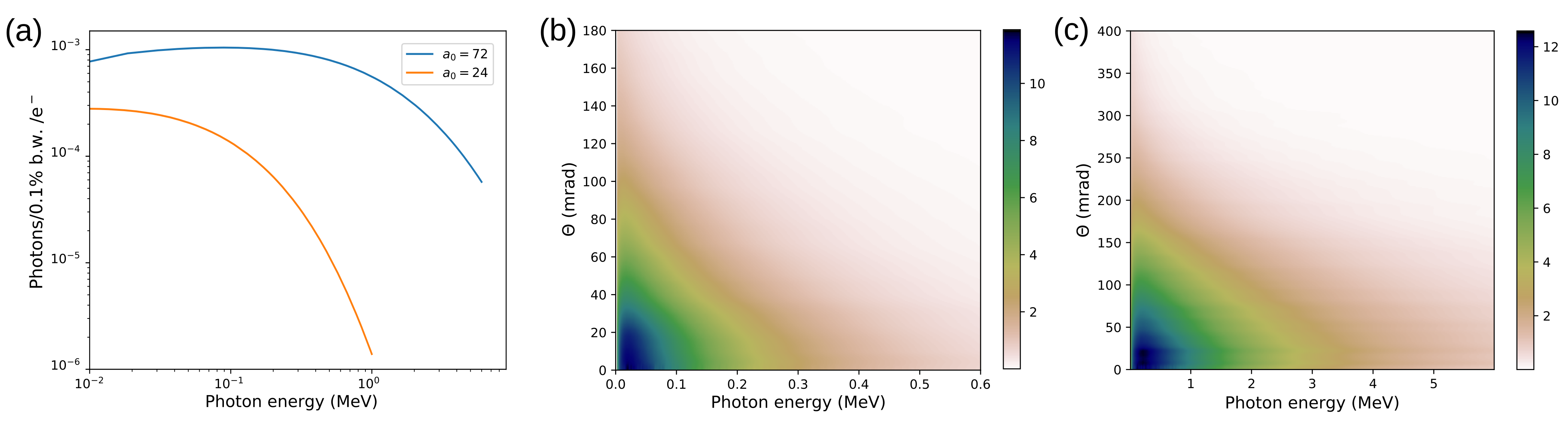}}
	\caption{ (a) Angular-integrated spectra of photon number calculated in 0.1\% bandwidth (BW).  The $\vartheta$-distribution of the betatron radiation spectra for $a_0 = 24$ (b) and $a_0 = 72$ (c).}
	\label{fig5}
\end{figure*}

The angular-integrated brightness of the x-ray radiation source is shown for different $a_0$ in \cref{fig5}a. The spectral brightness increases with both the laser intensity and $n_e$ increase. Similar behaves the critical frequency, $\omega_{crit}$, at which the spectrum sharply falls. The latter can be estimated with a good accuracy by the following expression \cite{Corde_2013}:
\begin{equation}
 E_{\rm crit} = \hbar\omega_{crit} \simeq 5\times 10^{-24}\gamma_0^2(n_{\rm e}/{\rm cm}^{-3})(r_{\beta 0}/\mu {\rm m}) {\rm keV}, \label{eq4}
\end{equation}
where $\gamma_0$ and $r_{\beta 0}$ are the Lorentz factor and the betatron amplitude, both averaged over the energies of the electrons forming a plateau in their spectra (at 50-200~MeV for $a_0 = 24$ and 100-400~MeV for $a_0 = 72$). For $a_0 = 24$ and $n_e = 0.1n_{cr}$, the values of these parameters are $\gamma_0\approx$ 300 and $r_{\beta 0}\approx 2$~$\mu$m, respectively, and, in accordance with \cref{eq4}, $E_{\rm crit}\approx$  0.15~MeV, while for $a_0 = 72$ and $n_e = 0.3n_{cr}$ they are $\gamma_0 \approx 500$ and $r_{\beta 0} \approx 2$~$\mu$m, and $E_{\rm crit}\approx 0.9$~MeV. The photon energies considerably exceed the ones achieved in the standard LWFA regime, e.g. \cite{Albert_2016}.  Note, that the betatron amplitude is roughly proportional to the bubble radius (which is somewhat larger), i.e. 
proportional to the ratio $\sqrt{a_0/n_e}$ (see \cref{eq1}), that is the same for both considered laser field amplitudes. This is why the $r_{\beta 0}$-values are the same. Since the critical frequency scales well with $\gamma_0^2 n_{\rm e}$, the latter can be used for its estimation for arbitrary parameters of the relativistic self-trapping regime. \Cref{fig5}b,c show the $\vartheta$-angular distributions of the betatron spectra, where $\vartheta = 0$ corresponds to the direction of laser pulse propagation. As usual, the most energetic photons are emitted in forward direction. The typical angular spread of the spectrum, is approximately 40~mrad for $a_0 = 24$ and 100~mrad for $a_0 = 72$, that is significantly lower than in the DLA regime, e.g. \cite{Huang_2016}. 

By using the electron spectra, which correspond to 7~nC ($a_0 = 24$) and 27~nC ($a_0 = 72$) for the particles with energies in exccess of 30~MeV we estimate the peak brightness of the synchrotron radiation given that the hard x-rays are emitted during approximately 75~fs (see \cref{fig4}) from the spot of  the size $\approx 4$~$\mu$m (see \cref{fig1}). This gives the following estimates for the peak brilliance as $1.2\times 10^{22}$~ph/s/mm$^2$/mrad$^2$/0.1\%b.w. at $\hbar\omega =100$~keV ($a_0 = 24$) and  $3\times 10^{22}$~ph/s/mm$^2$/mrad$^2$/0.1\%b.w. at $\hbar\omega =1$~MeV ($a_0 = 72$), that correresponds to state-of-art brilliance of the betatron source from the standard homogeneous plasmas (in contrast to more complicated schemes, when the brilliance can be substantially higher \cite{Ferri_2018}). The average number of emitted photons with energy exceeding 10~keV is $N_{\rm ph}\approx 0.64$ and $N_{\rm ph} \approx 5$ per one electron for  $a_0=24$ and $a_0=72$ that gives the total number of these photons $3\times 10^{10}$ and $8.4\times10^{11}$, respectively. This provides the conversion efficiency of $\simeq$ 0.01\% (4~J laser pulse) and $\simeq$ 0.02\% (36~J laser pulse), while the conversion efficiency to the entire keV-range of x-rays is estimated to be a few tenths of a percent.

In summary, we have demonstrated that in the regime of relativistically-intense laser pulse self-trapping in a near-critical density plasma the LWFA electrons with very high total charge emit a large amount of betatron x-ray radiation with the photon energy quanta in the range of 10~keV - 1~MeV, that is unprecedentedly high as compared to usual LWFA regime. Here, the effective loading of high number of electrons and stable soliton-like acceleration structure triggered by the high intensity laser pulse provides both strong accelerating plasma field and rather long acceleration length in a dense gas plasma. So, the proposed scheme of betatron x-ray generation benefits of both the cavity filling by laser field and the enhanced LWFA stage. We believe that experiments encouraged by such scheme are very attractive and timely. 

This work was supported by the Russian Science Foundation (grant no. 17-12-01283).

\end{document}